\documentstyle[12pt]{article}

\begin{document} 
\title{Spin-Torsion in Chaotic Inflation} 
\author{L. C. Garcia de Andrade\thanks{E-mail: garcia@dft.if.uerj.br} $\;$
and Rudnei O. Ramos\thanks{E-mail: rudnei@dft.if.uerj.br} \\ 
{\it Universidade do Estado do Rio de Janeiro, }\\ 
{\it Instituto de F\'{\i}sica - Departamento de F\'{\i}sica Te\'orica,}\\ 
{\it 20550-013 Rio de Janeiro, RJ, Brazil}} 
\date{July 1999}
\maketitle 
\thispagestyle{empty} 
\begin{abstract} 
\baselineskip 24pt 
The role of spin-torsion coupling to gravity is analyzed in the
context of a model of chaotic inflation. The system of equations
constructed from the Einstein-Cartan and inflaton field equations 
are studied  and it is shown that spin-torsion interactions
are effective only at the very first e-folds of inflation,
becoming quickly negligible and, therefore, not affecting 
the standard inflationary scenario nor the density perturbations 
spectrum predictions.
\vspace{0.34cm} 
\noindent 
PACS number(s): 98.80 Cq 
\end{abstract}
\newpage 
\setcounter{page}{1} 
Inflation, in its many different implementations,  has become one of
the most important cosmological paradigm today [for reviews, see
for instance, \cite{revinflat}]. The underlying idea of
inflation, of a period of accelerated expansion of the
scale factor, when the energy density is dominated by a vacuum energy 
density, is able to provide in a simple way a solution to
the cosmological horizon and flatness problems and at the same time
provides a model for density perturbations in the early Universe.
Earlier studies by Gasperini \cite{gasperini} have shown that inflation
could be driven by a spin density dominated epoch in the early 
Universe, even in the absence of vacuum dominant contributions to the
energy density, showing that a spin-torsion interaction acts like a
source of repulsive gravity. This then poses us with the question whether  
primordial spin-torsion interactions are able to support inflation
in standard inflaton driven inflationary scenarios, by, e.g., easing
the conditions for slow-roll of the inflaton field. 
Previous works on spin/torsion effects in inflation that we are aware
of \cite{torsion+inflat} have not detailed or elucidated
the real role of spin-torsion in an inflationary epoch. Torsion
makes an important role in very different physical models 
\cite{shapiro,sivaram}. In particular, torsion is natural to many models
of higher dimensional theory, as in Einstein-Kalb-Ramond models 
and string theory \cite{witten} and in gauge theories of
the Poincar\`e group \cite{group}. Therefore, it is natural to expect
that torsion may be particularly important in pre-inflationary models,
where quantum gravity effects may be introduced, from the geometrical
aspects of the space-time, by a torsion interaction term. This may
be the case in chaotic inflationary scenarios, where the inflaton 
initial conditions are taken around the Planck era and, then, quantum
gravity effects may become important to determine the initial
conditions prior to inflation.   
Based on the above motivations, in this letter, by considering
the spin-spin interactions of matter as described 
by the Einstein-Cartan theory (see, {\it e.g.}, Ref. \cite{group}), 
we study the role of spin-torsion
in the simplest model of chaotic inflation, which is that of 
an inflaton with a quadratic potential. We do not expect that
more general models of chaotic inflation will lead to results 
much different to this simple model, when regarding the effects
of spin-torsion, which is introduced through a generalization
of the gravity equations. 
In the Einstein-Cartan theory, the gravity equations are modified
such that the Friedman equation (we are assuming a spatially flat
Friedman-Robertson-Walker metric) and the acceleration
equations read \cite{gasperini}, respectively,
\begin{equation} 
H^2 = \frac{8 \pi G}{3} (\rho_\phi + \rho_{\rm s}) 
\label{G1} 
\end{equation} 
\noindent
and
\begin{equation} 
\frac{\ddot{a}}{a} = - \frac{4 \pi G}{3}  (\rho_\phi + 3 p_\phi -
8 \pi G \rho_{\rm s}) \;, 
\label{G2} 
\end{equation} 
\noindent
where  $H= \dot{a}/a$ is the Hubble parameter and $G= 1/M_{\rm
pl}^2$, with $M_{\rm pl}$ the Planck mass. In the above equations
we have also defined  $\rho_{\rm s}$, 
as $\rho_{\rm s} = \langle S_{\mu \nu \alpha} S^{\mu \nu \alpha}
\rangle/2$, the average of the square of the spin density tensor
$S_{\mu \nu \alpha}$. The spins are taken as randomly oriented 
(from not polarized spinning matter fields) \cite{gasperini},
so the average value of $S$ is zero. The torsion tensor
$Q_{\mu \nu}^{\;\;\alpha}$ is related with $S_{\mu \nu}^{\;\; \alpha}$
by the standard expression \cite{group}
\begin{equation}
Q_{\mu \nu}^{\;\;\alpha} = 8 \pi G \left( S_{\mu \nu}^{\;\; \alpha} +
\frac{1}{2} \delta_{\mu}^{\alpha} S_{\nu \beta}^{\;\; \beta}
- \frac{1}{2} \delta_{\nu}^{\alpha} S_{\mu \beta}^{\;\; \beta} \right) \;.
\label{torsion}
\end{equation}
\noindent
In Eqs. (\ref{G1}) and (\ref{G2}), $\rho_{\phi}$ and
$p_{\phi}$ are the energy density and pressure for the inflaton field
$\phi$, respectively, given by the usual relations:
$\rho_\phi = \frac{1}{2} \dot{\phi}^2 + V(\phi)$, $p_\phi = \frac{1}{2} 
\dot{\phi}^2 - V(\phi)$. As the spin-torsion does not couple to the 
inflaton field, we have the evolution equation for $\phi$,
\begin{equation} 
\ddot{\phi} + 3 H \dot{\phi} + V'(\phi) = 0 \;, 
\label{field} 
\end{equation}
\noindent
where in the above equations overdot represent derivative with respect 
to time and $V' = \frac{dV}{d \phi}$ is the field derivative of the
inflaton potential.
{}Using the simplest inflaton potential for chaotic inflation,
$V(\phi)= m^2 \phi^2/2$, and from Eqs. (\ref{G1}), (\ref{G2}) and
(\ref{field}), we determine the system of equations satisfying the
scale factor $a$, $\phi$ and $\rho_{\rm s}$,
\begin{eqnarray}
\ddot{a} &=& -\frac{8 \pi G}{3} \, a \, \left( \dot{\phi}^2 -
\frac{m^2}{2} \phi^2 -4 \pi G \rho_{\rm s} \right)  
\nonumber \\
\ddot{\phi} &=& -3 \frac{\dot{a}}{a} \dot{\phi} - m^2 \phi
\label{sys} \\
\dot{\rho_{\rm s}} &=& - 6 \frac{\dot{a}}{a} \rho_{\rm s}
\;. \nonumber
\end{eqnarray}
{}From the equation for the acceleration in the set of
equations in (\ref{sys}), we can immediately see that a
spin-torsion density works in favor of inflation, when the
slow-roll conditions for the inflaton fields applies. 
However, under these circumstances, of a regime of inflation,
the Universe quickly enters in a de Sitter phase, with
$H \sim {\rm const.}$ and, therefore, from the last of the 
equations in (\ref{sys}), $\rho_{\rm s}$ satisfies
\begin{equation}
\rho_{\rm s} \sim e^{-6 H t}   \;\;\;({\rm in \; de \; Sitter})
\label{rho_s}
\end{equation}
\noindent
decreasing with the sixth power of the inverse of the scale factor
during the de Sitter phase and then the spin density quickly vanishes 
as soon the Universe enters in an inflationary phase. Note that 
$\rho_{\rm s}$ decreases much faster than the inflaton density (and
even faster than radiation energy densities),
which goes with the third power of the inverse of the scale 
factor during the de Sitter phase. Therefore, we do not find a
spin-torsion dominated inflation over the inflaton in  
chaotic inflation models in general, with spin-torsion
interactions very fast becoming subdominat right after the
first e-folds of inflation. This is consistent with earlier
findings from Kao in \cite{torsion+inflat}, which working
in the context of torsion in the ten-dimensional Kalb-Ramond    
theory, concludes that the torsion field vanishes at the end of
the inflationary era. 
{}For the same reasons above, we do not expect any contribution
of spin-torsion interactions to density perturbations in 
chaotic inflationary scenarios, once the spin-torsion density
is depleted well before quantum fluctuations first cross the
horizon.
We can ask what happens when the initial value for
$\phi$ is much smaller than the usual
value needed in the model above in the absence of spin-torsion
effects, $\phi_i \sim 3.4 M_{\rm pl}$, as required for sufficiently 
inflation ($\sim 70 $ e-folds of inflation) in the model. It seems
that, from the first of the equations in 
(\ref{sys}), we could arrange a spin-torsion dominated epoch
over the inflaton field, for a sufficiently large initial
value for $\rho_{\rm s}$. However, this seems not to be
the case, since, from Eq. (\ref{G1}), it  imposes a
limit on the initial value for $\rho_{\rm s}$, as $2 \pi G \rho_{{\rm s},i}$
cannot be larger than $\rho_{\phi,i}$.
Also, very specific models as the one discussed by Gasperini in
\cite{gasperini}, shows that a spin-torsion dominated inflation, with
the physical requirements of large enough e-foldings of inflation,
can only be attained if we require a extreme fine-tuning for the 
spin density prescription used in there. 
\begin{center} 
{\large \bf Acknowledgements} 
\end{center}
\vspace{0.5cm} 
This work was partially supported by Conselho Nacional de
Desenvolvimento Cient\'{\i}fico e Tecnol\'ogico - CNPq (Brazil).

\end{document}